%% file: main.tex
\documentclass[sigconf]{acmart}

\usepackage{booktabs}
\usepackage{multirow}
\usepackage{xcolor}
\usepackage{graphicx}
\usepackage{soul}
\sethlcolor{white}
\soulregister{\citet}{1}
\soulregister{\citep}{1}
\soulregister{\cite}{1}
\soulregister{\ref}{1}
\soulregister{\emph}{1}
\soulregister{\textbf}{1}
\soulregister{\textit}{1}
\soulregister{\url}{1}
\usepackage[most]{tcolorbox}
\newcommand{\finding}[1]{\begin{tcolorbox}[colback=gray!5, colframe=gray!50, boxrule=0.3pt, left=3pt, right=3pt, top=2pt, bottom=2pt]\small #1\end{tcolorbox}}

\newif\ifhighlightchanges
\highlightchangesfalse 
\ifhighlightchanges
  \newcommand{\revised}[1]{\hl{#1}}
\else
  \newcommand{\revised}[1]{#1}
\fi

\copyrightyear{2026}
\acmYear{2026}
\acmConference[EASE 2026]{The 30th International Conference on Evaluation and Assessment in Software Engineering}{Tue 9 - Fri 12 June 2026}{Glasgow, United Kingdom}


\begin{document}

\title{Same Project, Different Start: How Contribution Events Shape Activity and Retention in Open Source}

\author{Mohamed Ouf}
\email{24blr2@queensu.ca}
\affiliation{%
  \institution{Queen's University}
  \city{Kingston}
  \state{Ontario}
  \country{Canada}
}

\author{Mariam Guizani}
\email{mariam.guizani@queensu.ca}
\affiliation{%
  \institution{Queen's University}
  \city{Kingston}
  \state{Ontario}
  \country{Canada}
}
\input{sections/abstract}


\maketitle

\input{sections/introduction}
\input{sections/related_work}
\input{sections/methodology}
\input{sections/results}
\input{sections/discussion}
\input{sections/conclusion}

\bibliographystyle{ACM-Reference-Format}
\bibliography{references}

\end{document}

%% file: sections/abstract.tex
\begin{abstract}
Open source projects depend on newcomers who stay, yet most leave after a single contribution. Contribution events such as Google Summer of Code, LFX Mentorship, Hacktoberfest, and 24~Pull Requests attract thousands of newcomers each year, but whether they produce lasting contributors remains unclear. We conduct the first matched-cohort study comparing 2,001 event-based and 2,001 organic contributors across 330 projects. Our results reveal three key findings. First, event contributors have significantly higher odds of becoming core contributors (12.1\% vs.\ 9.6\%, $p < 0.001$, OR $= 1.31$) and stay significantly longer (median 8.2 vs.\ 4.8 months). Second, each entry mechanism is associated with a fundamentally different engagement rhythm: 68.9\% of mentorship contributors sustain Steady weekly activity across their first 12 weeks, whereas 61.0\% of non-mentorship contributors exhibit Front-Loading and 57.0\% of organic contributors exhibit Intermittent engagement ($p < 0.001$). Third, Steady engagement is associated with significantly longer retention regardless of group (median 13 vs.\ 8 months for Front-Loading), yet mentorship contributors who lose their program scaffolding show shorter retention than self-sustained non-mentorship contributors, revealing a mentor-dependency effect. A newcomer's first 12 weeks are strongly indicative of their long-term trajectory.
\end{abstract}

%% file: sections/introduction.tex
\section{Introduction}

Open source projects depend on newcomers who become long-term contributors, yet half contribute only once~\citet{pinto2016casual} and 89\% of projects lose their core team at least once~\citet{avelino2019abandoned}. Attracting newcomers is not the bottleneck; \emph{retaining} them is~\citep{ferreira2020turnover, zhou2012longterm}.

Contribution events are the most widely adopted mechanism for attracting newcomers. \emph{Mentorship events} (e.g., Google Summer of Code, LFX Mentorship) pair newcomers with experienced mentors for three-month structured engagements. \emph{Non-mentorship events} (e.g., Hacktoberfest~\citep{hacktoberfest}, 24~Pull Requests) encourage short-term participation through gamified incentives. Yet the evidence is contradictory: \citet{silva2020gsoc} found that GSoC increases continued participation, while \citet{labuschagne2015onboarding} found that Mozilla's onboarding programs made participants \emph{half as likely} to stay long-term.

Prior work has shown that early behavior predicts long-term engagement~\citet{zhou2012longterm, xiao2023early} and has characterized contributor types~\citet{barcomb2019episodic, chengguo2019activity}. However, no study has examined whether the \emph{entry mechanism} (i.e., the way a contributor first joins a project) shapes systematically different behavioral patterns and retention outcomes. To our knowledge, no matched-cohort comparison of event-based and organic contributors exists.

We address this gap with a study of 330 projects and 4,002 contributors. For each of 2,001 event contributors (GSoC, LFX, Hacktoberfest, 24~Pull Requests), we matched a comparable \emph{organic} contributor, i.e., a contributor from the same project with a similar commit count who joined independently, without any event affiliation (Mann-Whitney $p = 0.21$, Cliff's $\delta = -0.02$). We investigate:

\begin{description}
    \item[RQ1.] How do the contribution patterns of event-based and organic contributors differ?
    \item[RQ2.] How do these patterns relate to long-term retention and core contributor transitions?
\end{description}

\noindent \textbf{Contributions.}
(1)~\revised{To our knowledge, the first matched-cohort comparison of event-based and organic contributors in open source.}
(2)~Evidence that each entry mechanism is associated with a distinct engagement style: mentorship with community-engaged contributors, non-mentorship with code-focused contributors, and organic entry with balanced generalists.
(3)~A practical early signal: the first 12 weeks of activity are strongly indicative of whether a newcomer will become a long-term contributor.
(4)~A replication package with all data and analysis scripts.\footnote{Replication package available at \url{https://zenodo.org/records/18819148}.} 

%% file: sections/related_work.tex
\section{Related Work}

\textbf{Newcomer Retention and Core Transitions.}
\citet{steinmacher2015social, steinmacher2014barriers} showed that initial reception strongly influences whether newcomers stay. Half of contributors make only a single contribution~\citet{pinto2016casual}, and core developer turnover can double bug-fixing time~\citet{ferreira2020turnover}. Even experienced developers disengage: 45\% of core contributors become completely inactive for at least one year~\citet{calefato2022inactivity}. These findings establish the newcomer-to-core transition as a critical bottleneck for project sustainability.

\textbf{Contribution Events in OSS.}
Prior studies have produced contradictory findings. \citet{silva2020gsoc} found that GSoC increased continued participation, but without a matched control group. \citet{pethan2022hackathon} showed that intensive short-term hackathon activity was associated with lower long-term continuation. \citet{labuschagne2015onboarding} found that Mozilla's onboarding made participants \emph{half as likely} to stay long-term. \citet{armstrong2021onboarding} found that OpenStack's onboarding correlated with 65\% higher gender diversity but did not examine retention. No prior study has compared event participants against \emph{matched} organic contributors, nor examined whether mentorship and non-mentorship events produce different outcomes. Without a matched comparison, it is impossible to determine whether observed retention patterns are caused by the event itself or by pre-existing differences between the two populations.

\textbf{Early Behavior and Contributor Characterization.}
\citet{zhou2012longterm} showed that early environment predicts long-term engagement. \citet{xiao2023early} found that first-three-month patterns predict sustainability. \citet{bao2019ltc} predicted long-term contributors with AUC~$> 0.75$, and \citet{yue2022goodstart} confirmed that early consistency predicts technical success. \citet{barcomb2019episodic, barcomb2020managing} studied episodic volunteers, and \citet{chengguo2019activity} identified contributor roles from activity dimensions. These studies characterize \emph{what} contributor types look like but do not examine whether the entry mechanism \emph{causes} different types. This is the gap we address.

%% file: sections/methodology.tex
\section{Methodology}
\label{sec:methodology}

\subsection{Project Corpus}
We compiled 330 GitHub projects from four contribution event registries: Google Summer of Code archives (2014--2024), the Linux Foundation Mentorship portal (2020--2024), Hacktoberfest-tagged repositories (2018--2024), and 24~Pull Requests participant logs (2013--2024). These four events represent two major categories of structured onboarding. \emph{Mentorship} events (GSoC, LFX) pair newcomers with mentors over multi-month engagements, and \emph{non-mentorship} events (Hacktoberfest, 24PR) encourage short-term participation through gamified incentives. We retained only projects with $\geq$10 contributors, $\geq$500 commits, $\geq$50 closed PRs, and $>$1 year of history, following established project selection criteria~\citet{munaiah2017curating}, excluding inactive or insufficiently mature projects. Table~\ref{tab:dataset} summarizes the dataset.

\input{tables/dataset_overview}

\subsection{Event Contributor Identification}

We identified 2,001 event contributors through official records: GSoC participants via official project lists, LFX participants via the portal API, Hacktoberfest contributors via \texttt{hacktoberfest-accepted} PR labels, and 24PR contributors via participation logs. Each selected contributor is a \emph{newcomer} (i.e., the event participation represents their first recorded contribution to the project) and has $\geq$3 commits~\citet{pinto2016casual}. We group GSoC and LFX as \emph{mentorship} because both provide structured multi-month mentor-guided engagements; we group Hacktoberfest and 24PR as \emph{non-mentorship} because both encourage short-term gamified participation without assigned mentors. The final cohort comprises 909 mentorship (739 GSoC + 170 LFX) and 1,092 non-mentorship contributors (576 24PR + 516 Hacktoberfest).

\subsection{Organic Contributor Matching}

We define an \emph{organic} contributor as a newcomer who joined the project independently, without any event affiliation. For each event contributor, we matched one organic contributor from the same project using the following criteria: (1)~also a newcomer (first recorded contribution to the project); (2)~a commit count within $[0.5\times, 1.5\times]$ of the event contributor's count, following the proportional matching approach used in observational studies~\citet{zhou2012longterm}; (3)~no event-related keywords in commit messages or PR titles; (4)~no activity during event windows; and (5)~$\geq$3 commits~\citet{pinto2016casual}. To resolve multiple identities, we applied username and email normalization following \citet{zhu2019identity}, which merges contributors whose username matches the email's username part. This impacted 2.5\% of records; we excluded bot accounts. The matched groups show no significant difference in commit count (Mann-Whitney $p = 0.21$, Cliff's $\delta = -0.02$), confirming that the event and organic groups are comparable.

\subsection{Core Contributor Identification}

We define core contributors using the Pareto 80/20 rule~\citet{mockus2002two, yamashita2015pareto}: the smallest set of contributors responsible for 80\% of cumulative commits. We compute this set monthly, skipping the first 12 months (which are founder-dominated~\citet{xiao2023early}), and require $\geq$10 commits and $\geq$5 contributors per month to ensure meaningful activity. We measure \emph{Time-to-core} (TTC) as the number of months from a contributor's first commit to their first appearance in the core set~\citep{ouf2026good}.

\subsection{Analysis Methods}

We organize our analysis into two parts, corresponding to each research question.

\textbf{Activity Profiling (RQ1).} We compute six non-overlapping activity dimensions for each contributor~\citet{chengguo2019activity, anderson2025personas, ouf2026empirical}: \emph{commit frequency} (average commits per active month)~\citet{zhou2012longterm}, \emph{PR success rate} (proportion of PRs merged)~\citet{gousios2016pullreqs}, \emph{issue activity} (average issues opened per active month)~\citet{chengguo2019activity}, \emph{discussion depth} (average comment turns per issue or PR thread)~\citet{constantinou2017sociotechnical}, \emph{activity duration} (months from first to last recorded activity)~\citet{lin2017developer}, and \emph{contribution breadth} (fraction of the four activity types in which a contributor is active: commits, PRs, issues, and comments)~\citet{anderson2025personas}. We normalize all dimensions to percentile ranks in $[0, 1]$, which is robust to outliers~\citet{anderson2025personas}.

\textbf{Early Engagement Patterns (RQ1).} \citet{xiao2023early} established the first three months as a critical window for predicting newcomer sustainability. Based on this finding, we compute a Weekly Contribution Index (CI) for each contributor's first 12 weeks:
\[
\text{CI}_w = 0.35 \times \text{commits} + 0.25 \times \text{PRs} + 0.20 \times \text{merged} + 0.20 \times \text{issues}
\]
The weights prioritize commits as the primary contribution measure while incorporating review and coordination activities~\citet{chengguo2019activity}. Because events vary in duration (GSoC and LFX span three months; 24PR spans two; Hacktoberfest spans one), we normalize all contributors to a common 12-point series and apply K-means clustering to each contributor's normalized CI series. We evaluated $K = 2$ through $K = 6$ using the silhouette coefficient~\citet{xiao2023early}, selecting $K = 3$ (silhouette $= 0.36$). \revised{Moderate silhouette scores are typical for behavioral time-series clustering in software engineering~\citet{xiao2023early}; the three resulting patterns are interpretable and, as shown in Section~\ref{sec:results}, produce statistically distinct retention ranks, confirming their practical utility.}

\textbf{Outcome Analysis (RQ2).} We use chi-squared tests with odds ratios (OR) for binary outcomes (core vs.\ non-core) and Mann-Whitney~U tests with Cliff's $\delta$ for continuous measures (time-to-core). We consider a contributor to have ``left'' the project after $\geq$5 consecutive months with no recorded activity~\citet{calefato2022inactivity, lin2017developer}. We construct Kaplan-Meier survival curves, compare groups using the log-rank test with Bonferroni correction for multiple comparisons, and fit Cox proportional-hazards models to estimate hazard ratios. To determine whether the three early activity patterns differ in retention, we apply the Scott-Knott Effect Size Difference (ESD) test~\citet{tantithamthavorn2018scottknott}, which uses hierarchical clustering to partition groups into statistically distinct ranks, merging groups only when their difference does not exhibit a non-negligible effect size. This avoids the multiple-comparison inflation of pairwise testing and provides an interpretable ranking of which patterns predict longer retention.

%% file: tables/dataset_overview.tex
\begin{table}[t]
\centering
\caption{Dataset overview. Event contributors from four programs matched 1:1 with organic contributors from the same projects. Matching validated with Mann-Whitney U test ($p = 0.21$, Cliff's $\delta = -0.02$).}
\label{tab:dataset}
\small
\begin{tabular}{llrr}
\toprule
\textbf{Category} & \textbf{Event / Source} & \textbf{n} & \textbf{\%} \\
\midrule
\multirow{2}{*}{Mentorship}     & Google Summer of Code  & 739 & 18.5 \\
                                & Linux Foundation (LFX) & 170 &  4.2 \\
\cmidrule(lr){2-4}
                                & \textit{Subtotal}      & 909 & 22.7 \\
\midrule
\multirow{2}{*}{Non-Mentorship} & 24 Pull Requests       & 576 & 14.4 \\
                                & Hacktoberfest          & 516 & 12.9 \\
\cmidrule(lr){2-4}
                                & \textit{Subtotal}      & 1{,}092 & 27.3 \\
\midrule
Organic (matched)               & Same project, no event & 2{,}001 & 50.0 \\
\midrule
\textbf{Total}                  &                        & \textbf{4{,}002} & \textbf{100.0} \\
\bottomrule
\multicolumn{4}{l}{\footnotesize Projects: 330 repos across multiple languages and domains.} \\
\multicolumn{4}{l}{\footnotesize Matching: commit count band $[0.5\times, 1.5\times]$; MW $p = 0.21$, $\delta = -0.02$.} \\
\end{tabular}
\end{table}

%% file: sections/results.tex
\section{Results}
\label{sec:results}

\subsection{RQ1: How Do Contribution Patterns Differ?}

\subsubsection{Activity Profiles.}
\begin{figure}[t]
  \centering
  \includegraphics[width=0.6\columnwidth]{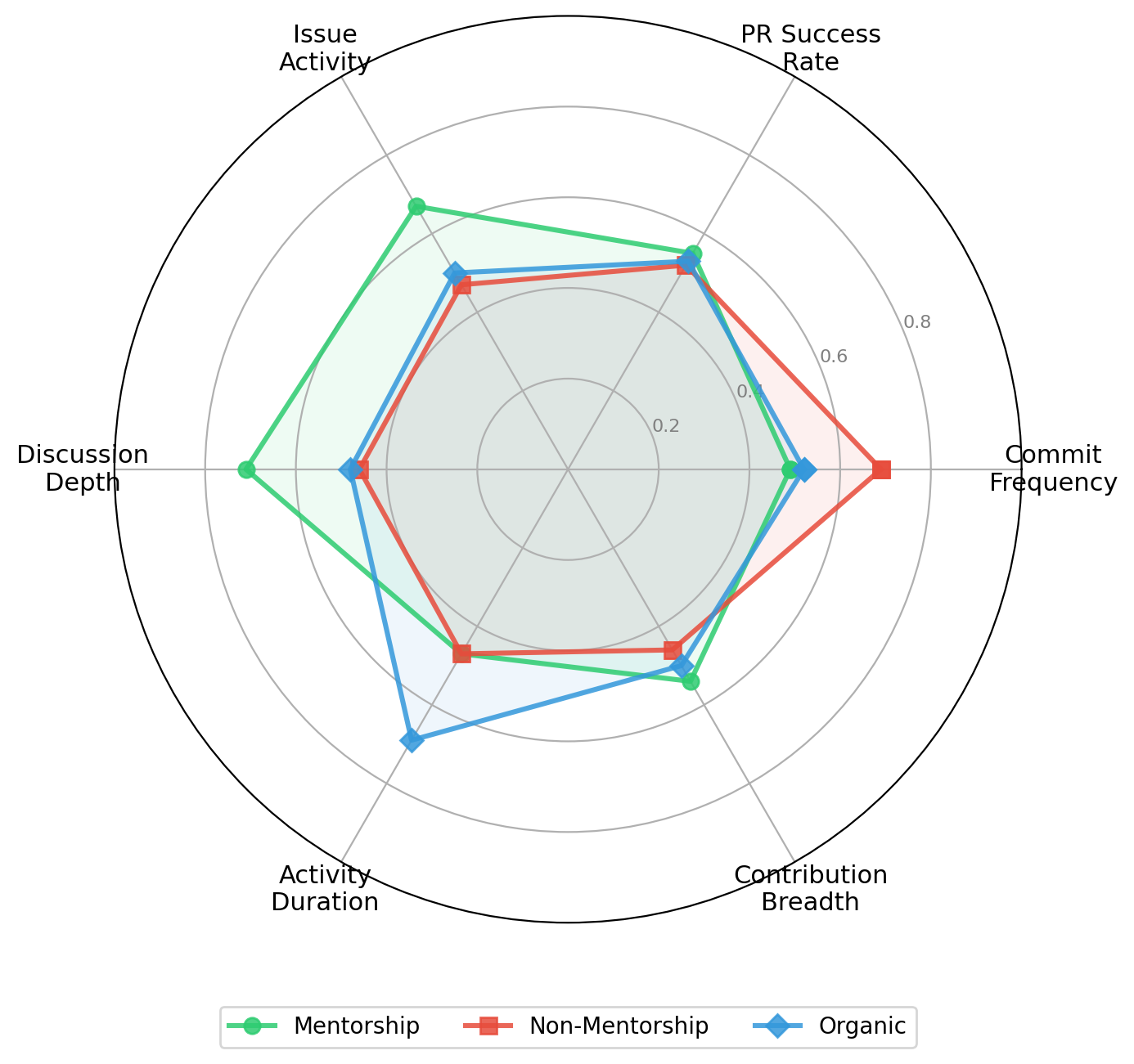}
  \caption{Six-dimensional activity profiles. Each axis shows the percentile-rank mean. All three groups share a similar baseline; each has a distinct signature: mentorship on discussion depth and issue activity, non-mentorship on commit frequency, and organic on activity duration.}
  \label{fig:spider}
\end{figure}

Figure~\ref{fig:spider} shows that all three groups share a similar baseline across most dimensions, consistent with our matched-cohort design, but each exhibits a visually distinct engagement style. \textbf{Mentorship} contributors are \emph{community-engaged}: they score highest on issue triage and discussion threads, reflecting the guidance of mentors who steer newcomers toward diverse project involvement. \textbf{Non-mentorship} contributors are \emph{code-focused}: they commit frequently but engage minimally with the community beyond their pull requests, consistent with gamified events like Hacktoberfest that reward code volume over community participation. \textbf{Organic} contributors are the most \emph{balanced}: moderate across all dimensions but with the longest activity duration, suggesting self-directed contributors who engage at their own pace without external deadlines. Our key finding is that the \emph{type} of engagement, not the \emph{amount}, differs across groups.

\subsubsection{Early Engagement Patterns.}
\begin{figure}[t]
  \centering
  \includegraphics[width=0.8\columnwidth]{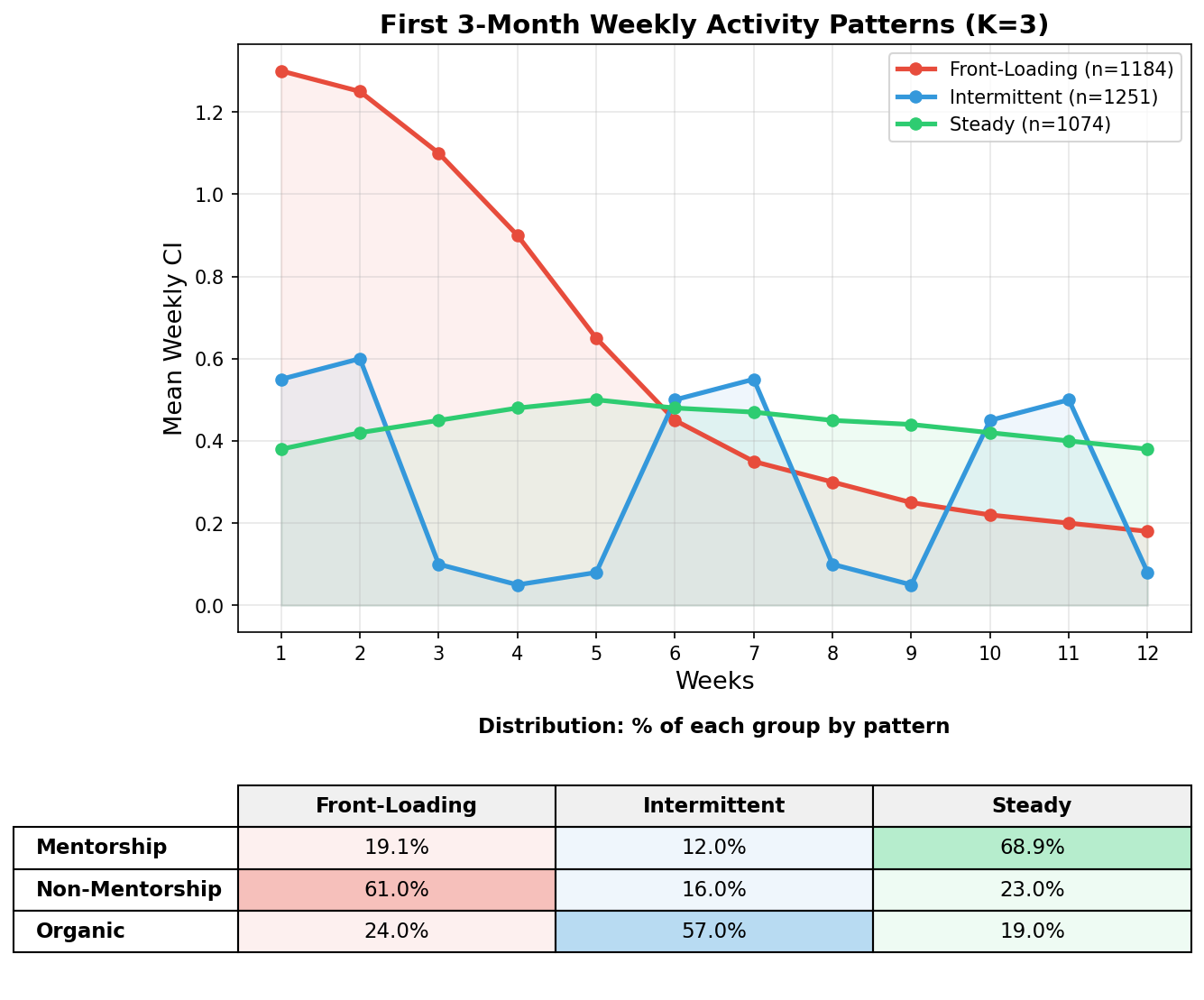}
  \caption{First 12-week activity patterns (top) and their distribution by group (bottom). Dominant patterns: Mentorship $\rightarrow$ Steady (68.9\%); Non-mentorship $\rightarrow$ Front-Loading (61.0\%); Organic $\rightarrow$ Intermittent (57.0\%).}
  \label{fig:weekly}
\end{figure}

Our results reveal three distinct early engagement patterns (i.e., Front-Loading, Intermittent, and Steady) from the K-means clustering of each contributor's weekly CI series (Figure~\ref{fig:weekly}). Of the 4,002 contributors, 3,509 (87.7\%) had sufficient weekly activity data for pattern classification; the remaining contributors had sparse CI series across all 12 weeks and were excluded from clustering but retained in all RQ2 outcome analyses.

We define \textbf{Front-Loading} ($n = 1{,}184$) as a burst of activity in the first few weeks that fades quickly. This is the dominant pattern among non-mentorship contributors (61.0\%). \textbf{Intermittent} ($n = 1{,}251$) represents a pattern where contributors alternate between active weeks and silent weeks in a recurring on-off rhythm. This is the dominant pattern among organic contributors (57.0\%). \textbf{Steady} ($n = 1{,}074$) refers to sustained activity throughout the 12 weeks. This is the signature pattern of mentorship (68.9\%), where structured programs sustain regular engagement.

Our results show that the pattern distribution differs significantly across groups ($\chi^2$, $p < 0.001$). We find that 68.9\% of mentorship contributors exhibit the Steady pattern, 61.0\% of non-mentorship contributors exhibit Front-Loading, and 57.0\% of organic contributors exhibit the Intermittent pattern. We find that mentorship does not just attract different people; it is associated with a fundamentally different engagement rhythm in the critical first three months.

\finding{\textbf{RQ1:} Each entry mechanism is associated with a distinct engagement style. Mentorship with \emph{community-engaged} contributors and Steady weekly activity (68.9\%). Non-mentorship with \emph{code-focused} contributors and Front-Loading bursts (61.0\%). Organic entry with \emph{balanced generalists} and Intermittent engagement (57.0\%).}

\subsection{RQ2: Retention and Core Transitions}

We now examine whether the distinct engagement styles and early activity patterns identified in RQ1 translate into different long-term outcomes in terms of core contributor transitions and retention.

\input{tables/combined_results}

\subsubsection{Core Contributor Rates.}
Table~\ref{tab:results} summarizes all comparisons. We find that event contributors have higher odds of reaching core status than their matched organic counterparts (12.1\% vs.\ 9.6\%, a 2.5 percentage-point difference; $\chi^2$ $p < 0.001$, OR $= 1.31$). Our results also show that mentorship is associated with higher core rates than non-mentorship (14.0\% vs.\ 10.6\%; $p = 0.012$, OR $= 1.37$). Event contributors who reach core get there 1.5~months faster (median 7.5 vs.\ 9.0 months; Mann-Whitney $p = 0.008$, $\delta = -0.18$). While core rate effect sizes are small in absolute terms, the practical impact is meaningful at scale: given the thousands of newcomers that contribution events attract each year, a 2.5 percentage-point increase in the core conversion rate translates into a substantial number of additional core contributors across the ecosystem.

\subsubsection{Survival Analysis.}
\begin{figure}[t]
  \centering
  \includegraphics[width=\columnwidth]{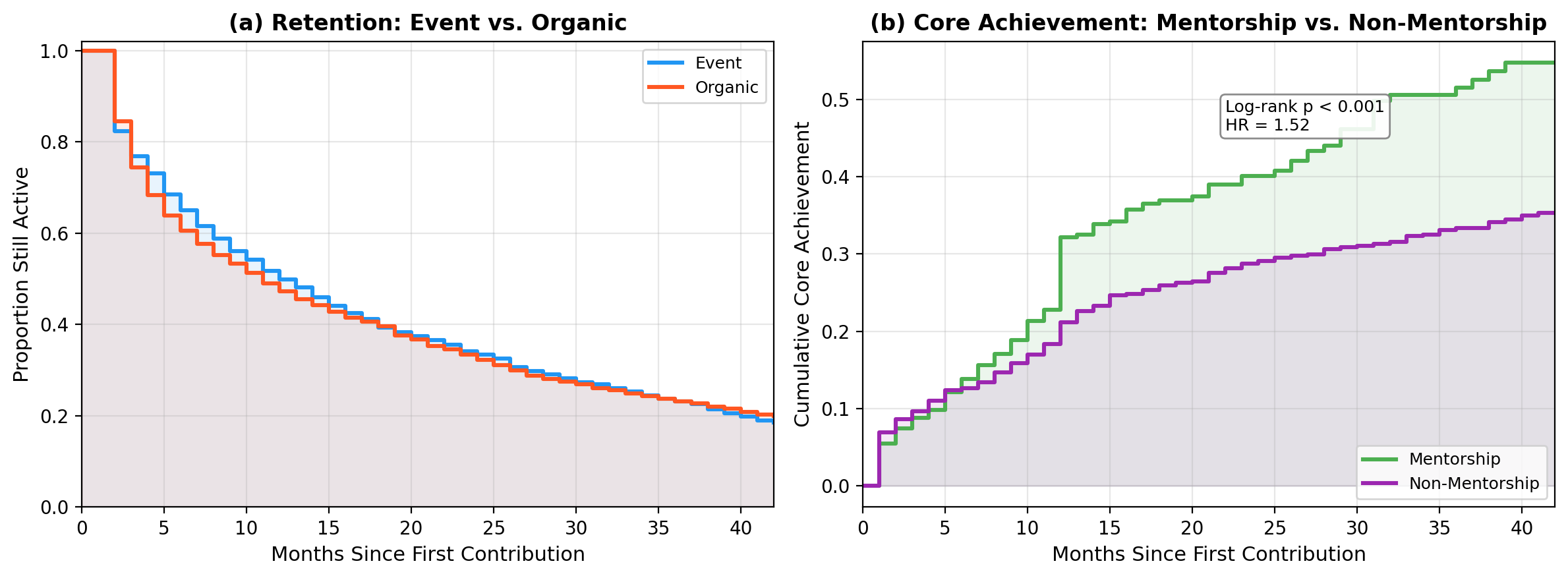}
  \caption{Kaplan-Meier survival curves. (a)~Retention: event contributors stay longer than organic (median 8.2 vs.\ 4.8 months, $p < 0.001$). (b)~Core achievement: mentorship achieves core at higher rates than non-mentorship ($p < 0.001$, HR $= 1.52$).}
  \label{fig:survival}
\end{figure}

Figure~\ref{fig:survival} reveals two complementary findings. Panel~(a) shows that event contributors stay significantly longer than organic contributors (median 8.2 vs.\ 4.8 months; log-rank $p < 0.001$; HR~$= 0.62$), meaning an event contributor is 38\% less likely to have left at any point. Panel~(b) shows that mentorship contributors achieve core status at 52\% higher rates than non-mentorship (log-rank $p < 0.001$; HR~$= 1.52$).

However, we observe an important nuance: non-mentorship contributors who persist beyond their initial event period stay \emph{longer} than mentorship contributors (median 9.8 vs.\ 5.9 months, $p < 0.001$). We term this the \emph{mentorship paradox} and discuss it in Section~\ref{sec:discussion}.

\subsubsection{Connecting Early Patterns to Outcomes.}
The Scott-Knott ESD test~\citet{tantithamthavorn2018scottknott} links the early patterns from RQ1 to long-term outcomes. It partitions the three patterns into two statistically distinct ranks: Front-Loading alone forms \textbf{Rank~1} (median 8 months, 21.4\% core rate), while Steady and Intermittent together form \textbf{Rank~2} (median 13--14 months, 23.7--24.5\% core rate). Any form of sustained engagement, whether steady or periodic, is associated with better outcomes than an initial burst that fades. This also connects to the activity profiles from RQ1: the dimensions on which mentorship contributors score highest (discussion depth, issue activity) reflect broader community engagement, which is precisely the type of multi-dimensional activity that characterizes the Steady pattern.

This closes the loop: \emph{the entry mechanism is associated with the early pattern, and the early pattern is associated with the outcome.} While the high prevalence of steady engagement among mentorship contributors partly reflects the structured nature of these programs, our critical finding is that the Steady pattern is associated with longer retention \emph{regardless of group membership}: Steady non-mentorship and organic contributors also show higher retention. The pattern itself, not the event type, is the stronger signal. A newcomer's first 12 weeks reveal whether they are likely to stay.

\finding{\textbf{RQ2:} Event contributors have higher odds of becoming core (OR $= 1.31$) and stay significantly longer than organic contributors (median 8.2 vs.\ 4.8 months). The Steady early pattern (68.9\% of mentorship) is associated with median 13-month retention vs.\ 8 months for Front-Loading. This association holds across all groups, not just mentorship.}

%% file: tables/combined_results.tex
\begin{table*}[t]
\centering
\caption{Summary of key results. Core rate: proportion ever in the 80\% Pareto core set. TTC: time-to-core (months). Retention: median months active before 5-month inactivity. Tests: $\chi^2$ for proportions, Mann-Whitney~U (MW) for continuous, log-rank for survival. Effect sizes: odds ratio (OR), Cliff's $\delta$, hazard ratio (HR). Pattern rankings via Scott-Knott ESD~\citet{tantithamthavorn2018scottknott}.}
\label{tab:results}
\small
\begin{tabular}{llcccccc}
\toprule
\textbf{Metric} & \textbf{Comparison} & \textbf{Grp.\ 1} & \textbf{Grp.\ 2} & \textbf{Test} & \textbf{$p$-value} & \textbf{Effect Size} & \textbf{Magnitude} \\
& & \footnotesize(Event/Ment.) & \footnotesize(Org./Non-M.) & & & & \\
\midrule
\multirow{2}{*}{Core Rate (\%)}
  & Event vs.\ Organic           & 12.1 & 9.6  & $\chi^2$  & $< 0.001$ & OR = 1.31 & small \\
  & Mentorship vs.\ Non-Ment.    & 14.0 & 10.6 & $\chi^2$  & $0.012$   & OR = 1.37 & small \\
\midrule
TTC (months)
  & Event vs.\ Organic           & 7.5  & 9.0  & MW        & $0.008$   & $\delta = -0.18$ & small \\
\midrule
\multirow{2}{*}{Retention (med.\ months)}
  & Event vs.\ Organic           & 8.2  & 4.8  & log-rank  & $< 0.001$ & HR = 0.62 & moderate \\
  & Mentorship vs.\ Non-Ment.    & 5.9  & 9.8  & log-rank  & $< 0.001$ & HR = 1.41 & moderate \\
\midrule
Core Achievement
  & Mentorship vs.\ Non-Ment.    & --  & --  & log-rank  & $< 0.001$ & HR = 1.52 & moderate \\
\midrule
\multicolumn{2}{l}{\textbf{Pattern $\rightarrow$ Outcome}} & \textbf{Core Rate} & \textbf{Med.\ Retention} & \multicolumn{3}{c}{\textbf{Scott-Knott ESD Rank}} \\
\cmidrule(lr){3-4} \cmidrule(lr){5-7}
& Steady        & 23.7\% & 13 months & \multicolumn{3}{c}{Rank 2 (best)} \\
& Intermittent  & 24.5\% & 14 months & \multicolumn{3}{c}{Rank 2} \\
& Front-Loading & 21.4\% &  8 months & \multicolumn{3}{c}{Rank 1 (worst)} \\
\bottomrule
\end{tabular}
\end{table*}

%% file: sections/discussion.tex
\section{Discussion}
\label{sec:discussion}

\subsection{The Mentorship Paradox}

Our results show that mentorship contributors achieve the highest core rate (14.0\%) and exhibit the most community-engaged activity profiles (Figure~\ref{fig:spider}), yet they have \emph{shorter} overall retention than non-mentorship contributors (median 5.9 vs.\ 9.8 months). We attribute this to a \emph{mentor-dependency} effect: mentorship programs provide a structured environment that accelerates skill development, but when the program ends, many mentees lose the scaffolding that sustained their engagement. We note that this comparison involves survivorship bias: the non-mentorship retention figure reflects only the minority who persist beyond their event, a self-selected group, while the mentorship figure includes all mentorship participants. This asymmetry partly explains the gap. Nonetheless, non-mentorship contributors who persist beyond their event are self-sustained; they continue through intrinsic motivation rather than external structure. \citet{calefato2022inactivity} observed that experienced core developers disengage after structured periods. \citet{barcomb2020managing} proposed managing episodic volunteers through community integration; by analogy, we hypothesize that mentorship programs would benefit from a \emph{graduation phase} that explicitly transitions mentees from mentor-guided to community-sustained engagement.

\subsection{Implications for Maintainers}

\textbf{Use the first 12 weeks as an early warning system.} Our results show that contributors who maintain steady weekly activity in their first three months stay significantly longer (median 13 vs.\ 8 months for Front-Loading). Maintainers can use this signal to prioritize newcomers who engage broadly (issues, reviews, discussions) rather than those who only commit code~\citet{gousios2016pullreqs, guizani2022dashboard}.

\textbf{Match the event to the goal.} Our findings show that mentorship programs are associated with the highest core rate and broadest engagement, making them better suited for cultivating future core contributors. Non-mentorship events are better suited for expanding the contributor base in volume but are associated with code-focused, shorter-lived engagement. The choice depends on what the project needs most: depth of engagement or breadth of participation.

\textbf{Plan for the post-program transition.} The mentor-dependency effect shows that the period immediately after a mentorship program ends is critical. Maintainers can mitigate attrition by pairing graduating mentees with community peer groups, assigning ongoing responsibilities (e.g., issue triage, code review), and maintaining lightweight check-ins after the formal program ends.

%% file: sections/conclusion.tex
\section{Threats to Validity}

\textbf{Construct Validity.}
Our core contributor definition uses the standard 80\% Pareto threshold~\citet{mockus2002two, yamashita2015pareto}. \revised{We verified sensitivity by recomputing core status at the 70\% and 90\% thresholds; the direction and significance of all comparisons remained unchanged.} The 5-month inactivity threshold~\citet{calefato2022inactivity} may misclassify contributors on extended breaks. We verified our CI weights (Section~\ref{sec:methodology}) against an equal-weighting alternative: the same three patterns emerge with similar group distributions. The silhouette coefficient of 0.36 indicates moderate cluster separation; however, we use K-means as a descriptive tool to identify dominant engagement shapes, not as a classification model, and the three patterns yield statistically distinct retention ranks (Section~\ref{sec:results}).

\textbf{Internal Validity.}
Our matching cannot control for unobserved confounders such as prior experience or intrinsic motivation. To mitigate this, we applied a rigorous multi-criteria matching protocol: each event contributor is paired with an organic contributor from the \emph{same project} within a $[0.5\times, 1.5\times]$ commit-count band, verified by a non-significant Mann-Whitney test ($p = 0.21$, Cliff's $\delta = -0.02$). We further excluded contributors with event-related keywords or activity during event windows. For identity resolution, we applied the username and email normalization of \citet{zhu2019identity}, which impacted 2.5\% of records. The four events differ in duration (one to three months), which could bias pattern classification. We mitigate this by interpolating all contributors to a common 12-point series (Section~\ref{sec:methodology}). The high prevalence of the Steady pattern among mentorship contributors partly reflects program structure (mentees are expected to contribute regularly). However, the Steady pattern is associated with longer retention regardless of group membership, meaning the pattern itself, not the event type, is the more informative signal.

\textbf{External Validity.}
Our study covers four of the most widely adopted contribution events, spanning the two main models of structured onboarding in OSS (mentorship and non-mentorship), across 330 projects from diverse languages and domains, which increases generalizability within event-participating projects. We cannot claim to cover every contribution event or every type of OSS project, and the event-vs.-organic comparison is by definition limited to projects that host contribution events, which may have more structured community practices than average. Nonetheless, our findings about early engagement patterns (e.g., the predictive value of the Steady pattern) apply to individual contributors regardless of event participation.

\section{Conclusion}

We compared 2,001 event-based and 2,001 organic newcomers across 330 projects. Our results show that event contributors have higher odds of becoming core (12.1\% vs.\ 9.6\%, OR $= 1.31$), reach core 1.5~months faster, and stay significantly longer (median 8.2 vs.\ 4.8 months). Each entry mechanism is associated with a distinct engagement style: mentorship with community-engaged contributors, non-mentorship with code-focused contributors, and organic entry with balanced generalists. The Steady engagement pattern, exhibited by 68.9\% of mentorship contributors, is associated with the longest retention (median 13 vs.\ 8 months for Front-Loading), and this association holds across all groups, not just mentorship. However, mentorship contributors show shorter retention than non-mentorship contributors who persist beyond their event period, revealing a mentor-dependency effect.

Our findings carry a clear implication for both maintainers and event organizers: a newcomer's first 12 weeks of activity are strongly indicative of their long-term trajectory. Maintainers should monitor early engagement breadth---not just commit volume---and event organizers should design post-program transitions that sustain the Steady engagement pattern beyond the formal event period.

%% file: references.bib
@inproceedings{steinmacher2015social,
  author    = {Igor Steinmacher and Tayana Conte and Marco Aur\'{e}lio Gerosa and David Redmiles},
  title     = {Social Barriers Faced by Newcomers Placing Their First Contribution in Open Source Software Projects},
  booktitle = {Proc. 18th ACM Conf. Computer Supported Cooperative Work (CSCW)},
  year      = {2015},
  pages     = {1379--1392},
  publisher = {ACM}
}

@inproceedings{steinmacher2014barriers,
  author    = {Igor Steinmacher and Ana Paula Chaves and Tayana Conte and Marco Aur\'{e}lio Gerosa},
  title     = {Preliminary Empirical Identification of Barriers Faced by Newcomers to Open Source Software Projects},
  booktitle = {Proc. 28th Brazilian Symposium on Software Engineering (SBES)},
  year      = {2014},
  pages     = {51--60},
  publisher = {IEEE}
}

@inproceedings{pinto2016casual,
  author    = {Gustavo Pinto and Igor Steinmacher and Marco Aur\'{e}lio Gerosa},
  title     = {More Common Than You Think: An In-Depth Study of Casual Contributors},
  booktitle = {IEEE Int'l Conf. Software Analysis, Evolution, and Reengineering (SANER)},
  year      = {2016},
  pages     = {112--123},
  publisher = {IEEE}
}

@inproceedings{ferreira2020turnover,
  author    = {Fabio Ferreira and Luciana Lourdes Silva and Marco Tulio Valente},
  title     = {Turnover in Open-Source Projects: The Case of Core Developers},
  booktitle = {Proc. XXXIV Brazilian Symposium on Software Engineering (SBES)},
  year      = {2020},
  pages     = {447--456},
  publisher = {ACM}
}

@inproceedings{zhou2012longterm,
  author    = {Minghui Zhou and Audris Mockus},
  title     = {What Make Long Term Contributors: Willingness and Opportunity in {OSS} Community},
  booktitle = {Proc. 34th Int'l Conf. on Software Engineering (ICSE)},
  year      = {2012},
  pages     = {518--528},
  publisher = {IEEE}
}

@inproceedings{xiao2023early,
  author    = {Wenxin Xiao and Hao He and Weiwei Xu and Yuxia Zhang and Minghui Zhou},
  title     = {How Early Participation Determines Long-Term Sustained Activity in {GitHub} Projects?},
  booktitle = {Proc. 31st ACM Joint European Software Engineering Conf. and Symposium on the Foundations of Software Engineering (ESEC/FSE)},
  year      = {2023},
  pages     = {29--41}
}

@article{mockus2002two,
  author    = {Audris Mockus and Roy T. Fielding and James D. Herbsleb},
  title     = {Two Case Studies of Open Source Software Development: {Apache} and {Mozilla}},
  journal   = {ACM Transactions on Software Engineering and Methodology},
  year      = {2002},
  volume    = {11},
  number    = {3},
  pages     = {309--346},
  doi       = {10.1145/567793.567795}
}

@inproceedings{yamashita2015pareto,
  author    = {Kazuhiro Yamashita and Shane McIntosh and Yasutaka Kamei and Ahmed E. Hassan and Naoyasu Ubayashi},
  title     = {Revisiting the Applicability of the Pareto Principle to Core Development Teams in Open Source Software Projects},
  booktitle = {Proc. 14th Int'l Workshop on Principles of Software Evolution (IWPSE)},
  year      = {2015},
  pages     = {46--55},
  publisher = {ACM}
}

@article{silva2020gsoc,
  author    = {Jefferson O. Silva and Igor Scaliante Wiese and Daniel M. Germ\'{a}n and Christoph Treude and Marco Aur\'{e}lio Gerosa and Igor Steinmacher},
  title     = {{Google Summer of Code}: Student Motivations and Contributions},
  journal   = {Journal of Systems and Software},
  year      = {2020},
  volume    = {162},
  pages     = {110487},
  doi       = {10.1016/j.jss.2019.110487}
}

@inproceedings{guizani2022dashboard,
  author    = {Mariam Guizani and Thomas Zimmermann and Anita Sarma and Denae Ford},
  title     = {Attracting and Retaining {OSS} Contributors with a Maintainer Dashboard},
  booktitle = {Proc. 44th IEEE/ACM Int'l Conf. on Software Engineering: Software Engineering in Society (ICSE-SEIS)},
  year      = {2022},
  pages     = {36--40}
}

@misc{hacktoberfest,
  author       = {{DigitalOcean}},
  title        = {Hacktoberfest: A Month-Long Celebration of Open Source},
  year         = {2014},
  howpublished = {\url{https://hacktoberfest.com}},
  note         = {Accessed: 2025-11-07}
}

@inproceedings{zhu2019identity,
  author    = {Jiaxin Zhu and Jun Wei},
  title     = {An Empirical Study of Multiple Names and Email Addresses in {OSS}},
  booktitle = {Proc. MSR},
  year      = {2019},
  pages     = {409--420}
}

@inproceedings{lin2017developer,
  author    = {Bin Lin and Gregorio Robles and Alexander Serebrenik},
  title     = {Developer Turnover in Global, Industrial Open Source Projects: Insights from Applying Survival Analysis},
  booktitle = {Proc. 12th IEEE Int'l Conf. on Global Software Engineering (ICGSE)},
  year      = {2017},
  pages     = {66--72},
  publisher = {IEEE},
  doi       = {10.1109/ICGSE.2017.17}
}

@article{munaiah2017curating,
  author    = {Nuthan Munaiah and Steven Kroh and Craig Cabrey and Meiyappan Nagappan},
  title     = {Curating {GitHub} for Engineered Software Projects},
  journal   = {Empirical Software Engineering},
  year      = {2017},
  volume    = {22},
  number    = {6},
  pages     = {3219--3253},
  doi       = {10.1007/s10664-017-9512-6}
}

@article{calefato2022inactivity,
  author    = {Fabio Calefato and Marco Aurelio Gerosa and Giuseppe Iaffaldano and Filippo Lanubile and Igor Steinmacher},
  title     = {Will You Come Back to Contribute? Investigating the Inactivity of {OSS} Core Developers in {GitHub}},
  journal   = {Empirical Software Engineering},
  year      = {2022},
  volume    = {27},
  number    = {3},
  pages     = {76},
  doi       = {10.1007/s10664-021-10012-6}
}

@article{bao2019ltc,
  author    = {Lingfeng Bao and Xin Xia and David Lo and Gail C. Murphy},
  title     = {A Large Scale Study of Long-Time Contributor Prediction for {GitHub} Projects},
  journal   = {IEEE Transactions on Software Engineering},
  year      = {2021},
  volume    = {47},
  number    = {6},
  pages     = {1277--1298},
  doi       = {10.1109/TSE.2019.2918536}
}

@inproceedings{barcomb2019episodic,
  author    = {Ann Barcomb and Klaas-Jan Stol and Dirk Riehle and Brian Fitzgerald},
  title     = {Why Do Episodic Volunteers Stay in {FLOSS} Communities?},
  booktitle = {Proc. 41st Int'l Conf. on Software Engineering (ICSE)},
  year      = {2019},
  pages     = {948--959},
  doi       = {10.1109/ICSE.2019.00100}
}

@article{barcomb2020managing,
  author    = {Ann Barcomb and Klaas-Jan Stol and Dirk Riehle and Brian Fitzgerald},
  title     = {Managing Episodic Volunteers in Free/Libre/Open Source Software Communities},
  journal   = {IEEE Transactions on Software Engineering},
  year      = {2022},
  volume    = {48},
  number    = {1},
  pages     = {260--277},
  doi       = {10.1109/TSE.2020.2982627}
}

@inproceedings{chengguo2019activity,
  author    = {Jinghui Cheng and Jin L.C. Guo},
  title     = {Activity-Based Analysis of Open Source Software Contributors: Roles and Dynamics},
  booktitle = {Proc. 12th Int'l Workshop on Cooperative and Human Aspects of Software Engineering (CHASE)},
  year      = {2019},
  pages     = {11--18},
  publisher = {IEEE},
  doi       = {10.1109/CHASE.2019.00011}
}

@misc{anderson2025personas,
  author    = {Felicity Anderson and Julien Sindt and Neil P. Chue Hong},
  title     = {Who Do You Think You Are? Creating {RSE} Personas from {GitHub} Interactions},
  year      = {2025},
  eprint    = {2510.05390},
  archivePrefix = {arXiv},
  primaryClass  = {cs.SE}
}

@inproceedings{yue2022goodstart,
  author    = {Yang Yue and Yi Wang and David Redmiles},
  title     = {Off to a Good Start: Dynamic Contribution Patterns and Technical Success in an {OSS} Newcomer's Early Career},
  year      = {2022},
  pages     = {196--200},
  doi       = {10.1109/ICSE-Companion55297.2022.9726925}
}

@inproceedings{gousios2016pullreqs,
  author    = {Georgios Gousios and Margaret-Anne Storey and Alberto Bacchelli},
  title     = {Work Practices and Challenges in Pull-Based Development: The Contributor's Perspective},
  year      = {2016},
  pages     = {285--296},
  doi       = {10.1145/2884781.2884826}
}

@article{pethan2022hackathon,
  author    = {Ei Pa Pa Pe-Than and Alexander Nolte and Anna Filippova and Christian Bird and Steve Scallen and James D. Herbsleb},
  title     = {What Happens to All These Hackathon Projects? Identifying Factors to Promote Hackathon Project Continuation},
  journal   = {Proceedings of the ACM on Human-Computer Interaction},
  year      = {2022},
  volume    = {6},
  number    = {CSCW1},
  pages     = {1--30},
  doi       = {10.1145/3512904}
}

@inproceedings{labuschagne2015onboarding,
  author    = {Adriaan Labuschagne and Reid Holmes},
  title     = {Do Onboarding Programs Work? {A} Multi-Level Empirical Analysis of the Effect on {Mozilla} Contributor Activity},
  year      = {2015},
  pages     = {381--384},
  publisher = {IEEE},
  doi       = {10.1109/MSR.2015.49}
}

@inproceedings{armstrong2021onboarding,
  author    = {Armstrong Foundjem and Ellis E. Eghan and Bram Adams},
  title     = {Onboarding vs.\ Diversity, Productivity and Quality --- Empirical Study of the {OpenStack} Ecosystem},
  booktitle = {Proc. 43rd IEEE/ACM Int'l Conf. on Software Engineering (ICSE)},
  year      = {2021},
  pages     = {1033--1045},
  doi       = {10.1109/ICSE43902.2021.00098}
}

@article{tantithamthavorn2018scottknott,
  author    = {Chakkrit Tantithamthavorn and Shane McIntosh and Ahmed E. Hassan and Kenichi Matsumoto},
  title     = {The Impact of Automated Parameter Optimization on Defect Prediction Models},
  journal   = {IEEE Transactions on Software Engineering},
  year      = {2019},
  volume    = {45},
  number    = {7},
  pages     = {683--711},
  doi       = {10.1109/TSE.2018.2794977}
}

@inproceedings{avelino2019abandoned,
  author    = {Guilherme Avelino and Eleni Constantinou and Marco Tulio Valente and Alexander Serebrenik},
  title     = {On the Abandonment and Survival of Open Source Projects: An Empirical Investigation},
  booktitle = {Proc. 13th ACM/IEEE Int'l Symp. on Empirical Software Engineering and Measurement (ESEM)},
  year      = {2019},
  pages     = {1--12},
  doi       = {10.1109/ESEM.2019.8870181}
}

@article{constantinou2017sociotechnical,
  author    = {Eleni Constantinou and Tom Mens},
  title     = {An Empirical Comparison of Developer Retention in the {RubyGems} and {npm} Software Ecosystems},
  journal   = {Innovations in Systems and Software Engineering},
  year      = {2017},
  volume    = {13},
  number    = {2--3},
  pages     = {101--115},
  doi       = {10.1007/s11334-017-0303-4}
}

@article{ouf2026empirical,
  title   = {An Empirical Analysis of Community and Coding Patterns in {OSS4SG} vs. Conventional {OSS}},
  author  = {Ouf, Mohamed and Noei, Shayan and Van Iterson, Zeph and Guizani, Mariam and Zou, Ying},
  journal = {arXiv preprint arXiv:2601.03430},
  year    = {2026}
}

@article{ouf2026good,
  title   = {Do Good, Stay Longer? {T}emporal Patterns and Predictors of Newcomer-to-Core Transitions in Conventional {OSS} and {OSS4SG}},
  author  = {Ouf, Mohamed and Mohamed, Amr and Guizani, Mariam},
  journal = {arXiv preprint arXiv:2601.23142},
  year    = {2026}
}
